\documentclass[11pt]{article}
\usepackage{ifthen}
\usepackage{times,mathptm,color}
\usepackage{graphicx}
\usepackage{amsthm, amsmath, amssymb, enumerate, hyperref}

\newcommand{\change}[1]{}

\newcommand{\width}{\mathrm{w}}

\theoremstyle{plain}
\newtheorem{theorem}{Theorem}[section]
\newtheorem{lemma}[theorem]{Lemma}

\newtheorem{proposition}[theorem]{Proposition}

\theoremstyle{definition}
\newtheorem{definition}[theorem]{Definition}

\theoremstyle{remark}

\newtheorem{claim}[theorem]{Claim}
\def\squareforqed{\hbox{\rlap{$\sqcap$}$\sqcup$}}
\def\qed{\ifmmode\squareforqed\else{\unskip\nobreak\hfil
\penalty50\hskip1em\null\nobreak\hfil\squareforqed
\parfillskip=0pt\finalhyphendemerits=0\endgraf}\fi}

\newenvironment{proofof}[1]{\begin{trivlist}%
\item[]{\flushleft\em Proof of #1. }}
{\qed\end{trivlist}}

\newcommand{\comments}[1]{}
\newcommand{\sepAuthor}{.5in}
\newcommand{\sepAbstract}{.5in}
\newcommand{\skipKeywords}{30pt}
\newcommand{\sepTitle}{2ex}

\long\def\mytitlepage#1#2#3#4{
        \thispagestyle{empty}
    \vspace*{\sepTitle}
        \begin{center}
        {\Large\bf #1}

        \vspace{\sepAuthor}
        #2\\
        \medskip

        \vspace{\sepAbstract}
        {\Large Abstract}
        \end{center}

        \noindent{#3}
        \vskip\skipKeywords

        \noindent{#4}
        \clearpage
        }
\begin{document}
\mytitlepage{ Constant-degree graph expansions that preserve
treewidth } {{\large Igor Markov
%\footnote{Supported in part by the US Air Force Research Laboratory, NSF 0448189 and 0205288.}
\ \  and\ \ Yaoyun Shi
%\footnote{Supported in part by NSF 0347078 and 0622033.}
}\\
\vspace{1ex}
Department of Electrical Engineering and Computer Science\\
The University of Michigan\\
Ann Arbor, MI 48109-2121, USA\\
E-mail: \{imarkov$,$shiyy\}@eecs.umich.edu}
{%abstract
 Many hard algorithmic problems dealing with graphs,
 circuits, formulas and constraints admit polynomial-time
 upper bounds if the underlying graph has small treewidth.
 The same problems often encourage reducing
 the maximal degree of vertices to simplify
 theoretical arguments or address practical concerns.
 Such degree reduction can be performed through
 a sequence of splittings of vertices, resulting
 in an {\em expansion} of the original graph. We observe that the treewidth
 of a graph may increase dramatically if the splittings are not performed
 carefully. In this context we address the following natural question: is it possible
 to reduce the maximum degree to a constant without substantially increasing the treewidth?

 Our work answers the above question affirmatively. We prove that any simple
 undirected graph $G=(V, E)$ admits an expansion $G'=(V', E')$ with the maximum degree $\le 3$
and $\textrm{treewidth}(G')\le \textrm{treewidth}(G)+1$.
Furthermore, such an expansion will have no
more than $2|E|+|V|$ vertices and $3|E|$ edges; it can be computed
efficiently from a tree-decomposition of $G$. We also construct a
family of examples for which the increase by $1$ in treewidth cannot
be avoided. }{{\bf Keywords}: Treewidth, graph expansion, constant
degree, ternary graph, algorithms}

\section{Introduction}

Given a graph $G$, its {\em treewidth} $\width(G)$ is a combinatorial parameter that measures
to what extent a graph differs from a tree. It is defined in terms
of {\em tree decompositions}, which are tree-like drawings of $G$
satisfying certain constraints. The width of a specific tree
decomposition represents the amount of clustering required to
subsume cycles so as to make the graph look like a tree, and
$\width(G)$ is defined as the smallest width over all possible tree
decompositions of $G$. Formal definitions are given in Section
\ref{sec:def}.

Since its introduction by several authors independently in the
1980's, the notion of treewidth has found numerous applications in
algorithm design. Many hard combinatorial problems, such as
Independent Set, Vertex Cover, SAT and \#SAT, admit algorithms whose
running time is $poly(n)\exp(\width(G))$, where $n$ is the input
size, and $G$ is the underlying graph structure. Thus, when
$\width(G)=O(\log n)$, such algorithms run in polynomial time.
For example, given a CIRCUIT-SAT instance size $n$ and a width-$w$
tree decomposition of the circuit graph $G$, the {\em bucket
elimination} algorithm~\cite{LauritzenS88, Dechter99} can be used to compute the
number of satisfying assignments in time $n^{O(1)}\exp(w)$.
Computing the optimal tree decomposition is
NP-hard~\cite{ArnborgCP87}. But fortunately the well known algorithm
by Robertson and Seymour~\cite{RSX} computes a tree decomposition of
width $O(\width(G))$ in time $|G|^{O(1)}\exp(\width(G))$. Making use
of this algorithm, the bucket elimination algorithm for SAT achieves
the same complexity. In practice, reasonably good tree
decompositions can be found by replacing the Robertson-Seymour
algorithm with heuristics, which removes a runtime bottleneck and
allows the entire algorithm to run fast on inputs of small
treewidth, e.g., in the case of probabilistic
inference~\cite{LauritzenS88}. The recent survey by
Bodlaender~\cite{Bodlaender06} outlines a number of other examples.

  In addition to treewidth, other graph parameters have also been heavily used
in applications to estimate and moderate complexity. A particular
parameter focal to this work is the maximum degree $\Delta(G)$ of a
graph. It is often desirable to reduce the maximum degree of an
input graph through a sequence of vertex splittings --- a
high-degree vertex $w$ is replaced by an edge connecting two new
vertices $u$ and $v$, and each neighbor of $w$ is assigned to be a
neighbor of either $u$ or $v$. Thus the original graph $G=(V,E)$ is
replaced by a graph $G'=(V',E')$ that is called an {\em expansion}
of $G$. The expansion process can be iterated, after which each
vertex $v\in G$ is replaced by a tree $T_v$ in $G'$, and each
original edge $uv\in E$ corresponds to an edge in $E'$ that connects
a pair of leaves in $T_u$ and $T_v$.

  Degree reduction arises in several unrelated algorithmic contexts,
  motivated by conceptual simplification, such as the reduction of SAT to 3-SAT,
  or application-specific concerns. For example, VLSI circuits use gates
  with limited fan-in to facilitate placement and routing in dense two-dimensional
  silicon wafers. Large AND, OR and XOR gates frequently occur in high-level
  descriptions of digital logic but are routinely broken down into trees
  of gates with bounded fan-in. Fan-out optimization is performed because
  fan-outs with high electrical capacitance lead to high circuit delay.
  They are split into trees using buffer (repeater) gates with small fan-out.
  To this end, research from Intel \cite{SMCK04} points out that the number of buffers
  in VLSI circuits is increasing with every technology generation and may exceed 50\%
  of all gates in several years. To consider fan-in and fan-out optimization of
  VLSI circuits in our graph-based framework, we represent each gate by two vertices
  linked by an edge --- one vertex connects all fan-ins, and the
  other connects all fan-outs. Performing degree minimization thus
  takes care of both cases.

  A simple procedure (the {\em symmetric expansion}) constructs for a graph $G$ an expansion $G'$ with $\Delta(G')\le 3$,
  but smaller $\Delta(G')$ cannot be guaranteed in general.
  This procedure replaces each vertex $u\in V$ with a path
  graph containing one vertex $u_v$ for each vertex $v$ adjacent to $u$ in $G$, and
  replaces each $uv\in E$ by the edge connecting $u_v$ to $u_v$.
  However, not all expansions are equally favorable. In VLSI circuit optimization,
  the choices of tree do not affect the circuit's functionality, but they may
  increase the treewidth which will complicate placement and routing.

  Treewidth also features prominently in computational logic and
  algorithms for constraint-satisfaction problems. In particular,
  {\em directional resolution} introduced by Davis and Putnam
  in 1960 for solving CNF-SAT works  particularly well on instances with low
  width \cite{DechterR94}. More general constraint satisfaction problems
  with limited treewidth admit polynomial-time algorithms for finding and counting
  solutions \cite{DechterP87,DechterKBE02}, and the same applies
  to the evaluation of Bayesian networks \cite{Dechter99}.
  Therefore, preserving treewidth is crucial when transforming constraint systems,
  e.g., when converting SAT to 3-SAT. This particular problem is
  the focus of \cite{ZD07}, which proposes a specific transformation
  to generate 3-SAT instances at most seven times larger, whose
  treewidth is increased by at most one.

%  Here we give two such transformations. A high-degree variable can be expanded
%  into (any) tree of low-degree variables, where every edge corresponds to an equality constraint.
%  In the case of CNF-SAT, it is also possible to reduce the length of clauses using
%  the following procedure. First transform the CNF-SAT instance into an AND/OR/NOT circuit
%  where (i) each clause is modeled by an OR gate with one input per literal, and (ii) the
%  outputs of these OR gates are connected one giant AND gate. Second, break up
%  these multi-input gates into an AND-tree and, OR-trees
%  respectively, where every gate has no more than two inputs.
%  Third, apply a well-known transformation from Circuit-SAT
%  to CNF-SAT by representing each gate with several CNF clauses.
%  While the specific tree topologies do not affect the correctness of this
%  construction, they may increase the treewidth of the resulting
%  CNF-SAT instance and make it more difficult to solve.

\begin{figure}[tbph]
\label{fig:blownup-dots}
\centering
\includegraphics[width=5in]{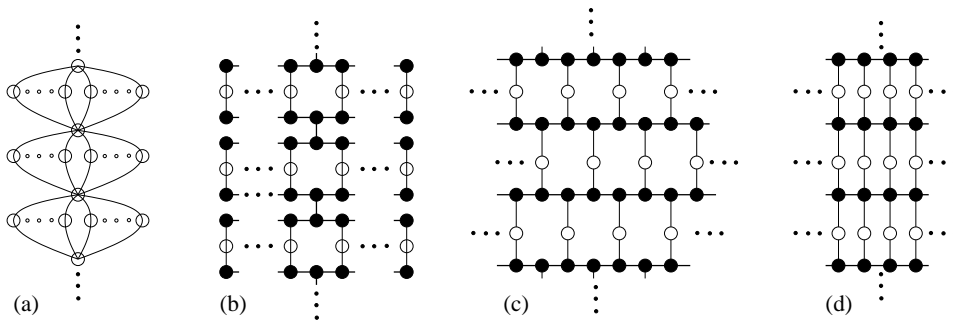}
\caption{The graph in (a) has treewidth 2, and has a ternary
expansion (b) of the same treewidth. A symmetric expansion, shown in
(c), however, contains a grid minor (d). Thus (c) has treewidth at
least $4$ and, if scaled, can reach an arbitrary treewidth.}
\end{figure}

  The considerations above lead to a natural question :
  is it possible to reduce the maximum degree of $G$ through vertex splitting
  without substantially increasing the treewidth of $G$?
  While treewidth cannot decrease during expansion, a simple example
  in Figure~\ref{fig:blownup-dots} illustrates that treewidth may dramatically
  increase after a symmetric expansion.

Let us call a graph $G$ {\em ternary} if $\Delta(G)\le 3$.
Note that the maximum degree of a ternary graph cannot be further
reduced by expansion.
Our main result gives an affirmative answer to the above question:
we prove that any graph $G=(V, E)$ admits a ternary expansion
with treewidth $\le \width(G)+1$.
We give a polynomial-time algorithm to compute
such an expansion from an optimal tree decomposition of $G$.
In general, given a tree decomposition of width $w$, the output of the algorithm
is a ternary expansion with width $\le w+1$.
Thus, combined with the Robertson-Seymour algorithm, our algorithm outputs
a ternary expansion $G'$ of $G$ with width $O(\width(G))$ in time $|G|^{O(1)}\exp(O(\width(G)))$.
Finally, we construct a family of graphs $G_n$
such that any ternary expansion $G'_n$ of $G_n$ must have $\width(G')=\width(G)+1$.
Thus our algorithm achieves
the generally possible minimum treewidth. Its additional
applications include a forthcoming work on quantum circuits where it
helps to establish an efficient classical algorithm for simulating a
broad class of quantum computations~\cite{MarkovS07}.

  The remaining part of the paper is structured as follows.
  We review the notions of tree decomposition and treewidth
  in Section \ref{sec:def}, then prove our main result in Section \ref{sec:main}.
  Section \ref{sec:increase} shows that our result cannot be improved,
  and final remarks are given in Section \ref{sec:discussion}.

\section{Definitions}
\label{sec:def}

Let $G=(V, E)$ be an undirected simple graph. For a vertex $v\in V$,
we denote the set of its adjacent vertices (neighbors) by $N(v)$.
Further $d(v)=|N(v)|$, and $\Delta(G)=\max_{v\in V} d(v)$.

Let $G$ be a graph. Following definitions in \cite{RSIII}, a {\em
tree decomposition} of $G$ is a tree $\mathcal{T}$, together with a
function that maps each tree vertex $w$ to a subset $B_w\subseteq
V(G)$ . These subsets $B_w$ are called {\em bags} and can be used as
vertex labels. In addition, the following conditions must hold.
\begin{enumerate}[(T1)]
\item $\bigcup_{v\in V(\mathcal{T})} B_v = V(G)$, i.e., each vertex
      must appear in some bag (and may appear in multiple bags).
\item $\forall \ \{ u, v\}\in E(G)$, $\exists w\in V(\mathcal{T})$,
$\{u, v\}\subseteq B_w$, i.e., for each edge, there must
be a bag containing both of its end vertices.
\item $\forall\ u\in V(G)$, the set of vertices $w\in V(\mathcal{T})$
with $u\in B_w$ form a connected subtree $T_u$, i.e., all bags containing
a given vertex must be connected in $\mathcal{T}$.
\end{enumerate}
The {\em width} of a tree decomposition $\mathcal{T}$, denoted by
$\width(\mathcal{T})$, is defined by $\max_{w\in V(\mathcal{T})}
|B_w|-1$. For graph $G$, its treewidth $\width(G)$ is the minimum
width of tree decompositions of $G$.
While NP-hard to compute in general, $\width(G)$ is known for common
classes of graphs~\cite{Diestel05} --- a non-empty tree has
treewidth $1$, the $n\times n$ grid has treewidth $n$, and a
parallel serial graph has treewidth $\le 2$.
Figure~\ref{fig:stage2} shows an example of a graph of treewidth $3$
and its tree decomposition of the same width.

A key motivation for the study of treewidth is the study of graph
minors. Let $G=(V, E)$ be a graph. The {\em contraction} of an edge
$uv\in E$  is the following operation on $G$: remove $u$ and $v$
(and all incident edges), and connect all neighbors of $u$ and $v$
to a new vertex $w$. A graph $G'$ is a minor of $G$ if $G'$ can be
obtained from a sequence of edge contractions on a subgraph of $G$.
In this case, a tree decomposition for $G$ also induces a tree
decomposition for $G'$ of equal or smaller width.
%  Thus $\width(G')\le \width(G)$.
Usually $\width(G')<\width(G)$, in particular, contracting all edges
of $G$ will reduce $G$ to an empty graph, which has treewidth $0$.

The process of {\em splitting} studied in our work can be viewed as
inverse to contraction.
\begin{definition}
Let $G=(V, E)$ be a graph. The {\em splitting} of $v\in V$ with the
support $S\subseteq N(v)$ is the following transformation of $G$:
introduce a new vertex $v'$, connect $v'$ to $v$, for any $s\in S$,
disconnect it from $v$ and connect it to $v'$. A graph $G'$
is called an {\em expansion} of $G$ if there exists a sequence of splittings
that transform $G$ to $G'$, and is said to be {\em irreducible} if
no degree-$2$ vertex is created in the splittings. If $\Delta(G')\le
3$, we call $G'$ a {\em ternary} expansion of $G$.
\end{definition}

It follows immediately from the definition of splitting a vertex $v$
that contracting the edge $vv'$ results in the original graph. Thus
if $G'$ is an expansion of $G$, then $G$ is a minor of $G'$, but not
{\em vice versa}, in general. It also follows from the definition
that $G'$ is an expansion of $G$ if and only if $G$ can be obtained
from $G'$ by contracting edges in a set of vertex-disjoint tree
subgraphs without creating any parallel edges. Furthermore, an
expansion $G'$ of $G$ is irreducible if and only if none of the
vertices involved in the contraction has degree $2$.
%As shown in the Introduction, $\tw(G)<\tw(G')$ is possible in this case.
We note that the size of any irreducible expansion must be linear in
the size of the original graph. Denote by $|V|_0$ the number of
degree-$0$ vertices in a graph $G=(V, E)$.
\begin{proposition}\label{prop:size}
Any irreducible expansion $G'=(V', E')$ of $G=(V, E)$ must have
$|V'|\le 2|E|+|V|_0$ and $|E'|\le 3|E|$.
\end{proposition}
\begin{proof}
Let $v\in V$ be a vertex split in creating $G'$. Then $d(v)\ge4$,
for otherwise a degree-$2$ vertex would be created, contradicting to
the assumption that $G'$ is irreducible. Denote by $T'_v$ the tree
subgraph of $G'$ whose internal vertices contract to $v$. Then the
number of leaves of $T'_v$ is precisely $d(v)$. Since $T'_v$ does
not have a degree-$2$ vertex, it follows from a simple induction
that $T'_v$ has $\le d(v)-2$ internal vertices and $\le d(v)-1$
internal edges. Therefore,
\[|V'|\le \sum_{v\in V: d(v)\ge 4} (d(v) -2)+ \sum_{v\in V: d(v)\le 3} 1 \le
\sum_{v\in V} d(v) + |V|_0 \le  2|E|+|V|_0,\]
and
\[ |E'| \le |E|+\sum_{v\in V: d(v)\ge 4} (d(v)-1) \le 3|E|.\]
\end{proof}

In our construction of an expansion, we may introduce degree-$2$
vertices for the convenience of bounding the treewidth. Such
vertices can be removed easily at the end to obtain an irreducible
expansion.

\section{Main result and its proof}
\label{sec:main}

\begin{theorem}
\label{th:main} There is a polynomial-time algorithm that, given a
graph $G=(V, E)$ and its tree decomposition of width $w$, computes
a ternary expansion $G'=(V', E')$ with $\width(G')\le w+1$.
In particular, $G$ admits a ternary expansion whose treewidth is no more than
$\width(G)+1$.
% and %$w=\width(G)$.
\end{theorem}

The construction of $G'$ takes several stages: first we construct
graph $G_1$ with a tree decomposition $\mathcal{T}_1$ such
that the subgraph of $G_1$ induced by each bag in $\mathcal{T}_1$
has maximum degree $\le 2$. In the second stage, each vertex $v$
is split many times to replicate the structure $T_v$, the tree formed by
those bags containing $v$ in $\mathcal{T}_1$.
Two vertex trees $T_u$ and $T_v$ for $uv\in E(G_1)$ are then connected
through a pair of vertices corresponding to the same bag that contains
$u$ and $v$. In the last stage, each vertex is split many times
to reduce the degree within its vertex tree. We combined
the last two stages in our following description of the construction.

\begin{proofof}{Theorem~\ref{th:main}}
If $\width(G)\le1$, then $G$ is a forest. Repeatedly splitting a
vertex with degree $\ge4$ with $2$ supporting vertices will result
in an expansion $G'=(V', E')$ with $\Delta(G')\le 3$, $|V'|\le
|V|+|E|=2|V|-1$, and $\width(G')\le1$. Thus the Theorem holds.
We now consider $\width(G)\ge2$.

\begin{figure}[tbph]
\centering
\includegraphics[width=2.5in]{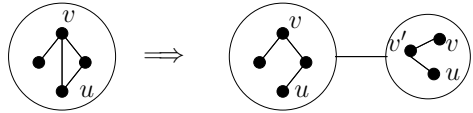}
\caption{Inside a bag, vertex $v$ of degree $3$ is split with the
support of a neighbor $u$ in the same bag.  A new neighboring bag is
created, containing $v$, $u$, and the new vertex $v'$ from the
splitting.} \label{fig:insidebag}
\end{figure}

{\bf Stage 1: Reducing the maximum degree inside a bag.} We
sequentially scan the bags of $\mathcal{T}$ and apply the following
operations. If a bag $B$ contains a vertex $v$ with $d(v)\ge 3$, let
$u\in B$ be a neighbor of $v$, split $v$ with the support $\{u\}$.
Denote the new vertex by $v'$. (This is equivalent to placing $v'$
at the edge $uv$.) Create a new bag $B'$ containing $\{u, v, v'\}$
and attach it to $B$. This process extends the $u$-subtree and the
$v$-subtree of $\mathcal{T}$ by a leaf bag $B'$, and adds a trivial
$v'$-subtree ($B'$),  thus results in a tree decomposition. Since
the new bag has $3$ vertices, and $\width(\mathcal{T})\ge 2$, the
width of the new tree decomposition remains the same.
Figure~\ref{fig:insidebag} illustrates the process of adding
one bag.

Denote by $G_1=(V_1, E_1)$ the resulting graph, and by
$\mathcal{T}_1=(\mathcal{V}_1, \mathcal{E}_1)$ the resulting tree
decomposition. Let $k\le |E|$ be the number of splittings. We have

\begin{enumerate}[(\thesection .a)]
\item \label{lb:degree}The maximum degree of an induced subgraph of $G_1$ by vertices in a bag
of $\mathcal{T}_1$ is at most $2$.
\item $\width(\mathcal{T}_1)=\width(\mathcal{T})$.
\end{enumerate}

{\bf Stage 2: Completing the construction.}
We now construct a graph $G'=(V', E')$ from $G_1$ and later show that it is
an expansion of $G_1$, thus an expansion of $G$.
Fix a total ordering $\preceq$ on $\mathcal{V}_1$.
Define $V'\subseteq V_1\times \mathcal{V}_1\times \mathcal{V}_1$ as follows
\[ V'=\{ (v, B, B'):
(v\in B\cap B')\wedge(B=B'\vee BB'\in \mathcal{E}_1)\}.\]

\begin{figure}[tbph]
\centering
\includegraphics[width=5in]{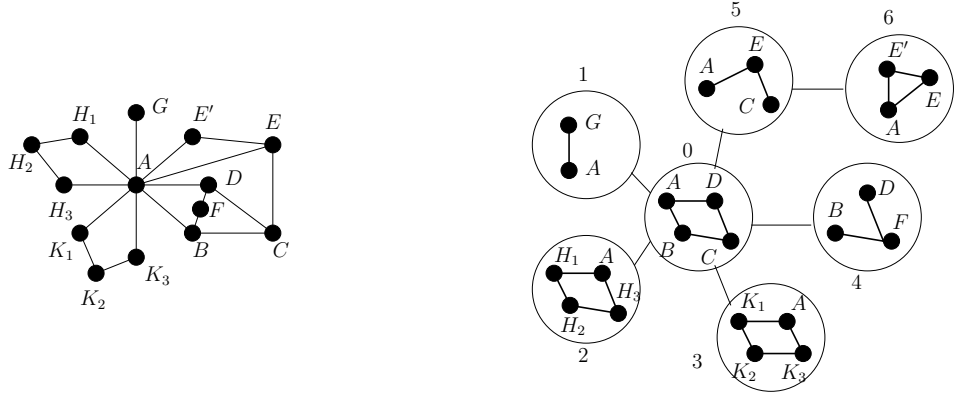}
\caption{A graph and a tree decomposition that satisfies the
condition in Stage 2 (of having $\le 2$ induced degree within each bag).}
\label{fig:stage2}
\end{figure}

For each $v\in V_1$, let $T_v$ be the $v$-subtree of $\mathcal{T}_1$.
There are three types of edges in $G'$.
\begin{enumerate}[(\thesection .i)]
\item Let $B_0$ be a bag, $v\in B_0$, and $B_1\prec B_2\prec\cdots\prec B_t$
are the neighbors of $B_0$ in $T_v$.
An {\em ordering edge} connects a vertex $(v, B_0, B_i)$ with $(v, B_0, B_{i+1})$, $0\le i\le t-1$.
\item An {\em intra-tree edge} connects $(v, B, B')$ with $(v, B', B)$, where $B\ne B'$,
and $BB'\in E(T_v)$. In this case, $v\in B\cap B'$.
\item An {\em inter-tree edge} connects $(v, B, B)$ with $(u, B, B)$, for $u\ne v$, $uv\subseteq E_1$.
In this case, $u, v\in B$.
\end{enumerate}
The edge set $E'$ consists of all possible ordering and intra-tree edges, but only one inter-tree edge for one edge
of $E_1$.
Figures~\ref{fig:stage2} and \ref{fig:stage2-exp} illustrate the operations in Stage 2.

Stages 1 and Stage 2 take polynomial time. Using the following
claims, we verify that $G'$ satisfies the properties in the Theorem.
\end{proofof}

\begin{claim} The graph $G'$
is an expansion of $G$.
\end{claim}
\begin{proof}
Contracting all ordering edges associated with an occurrence of $v$
in a bag $B_0$ combines all $(v, B_0)$ into a single vertex
associated with this occurrence. All those vertices are connected
through intra-tree edges, and this graph is precisely $T_v$. Since
$T_v$ and $T_u$ are connected through one inter-tree edge if and
only if $uv\in E$, contracting the intra-tree edges (after
contracting the ordering edges) gives $G$. Since no parallel edges
were created in the whole process, $G'$ must be an expansion of $G$.
\end{proof}

\begin{claim} The graph $G'$ satisfies $\Delta(G')\le 3$.
\end{claim}
\begin{proof}
A vertex $(u, B, B)$ is incident to at most one ordering edge, no
intra-tree edge, and at most two inter-tree edges. By Property
(\thesection .\ref{lb:degree}), its degree $\le 3$. A vertex $(u, B,
B')$ with $B\ne B'$ is incident to at most two ordering edges, at
most one intra-tree edge (connecting $(u, B', B)$), and no
inter-tree edge. Therefore its degree $\le 3$, and $\Delta(G')\le
3$.
\end{proof}

\begin{figure}[tbph]
\centering
\includegraphics[width=6in]{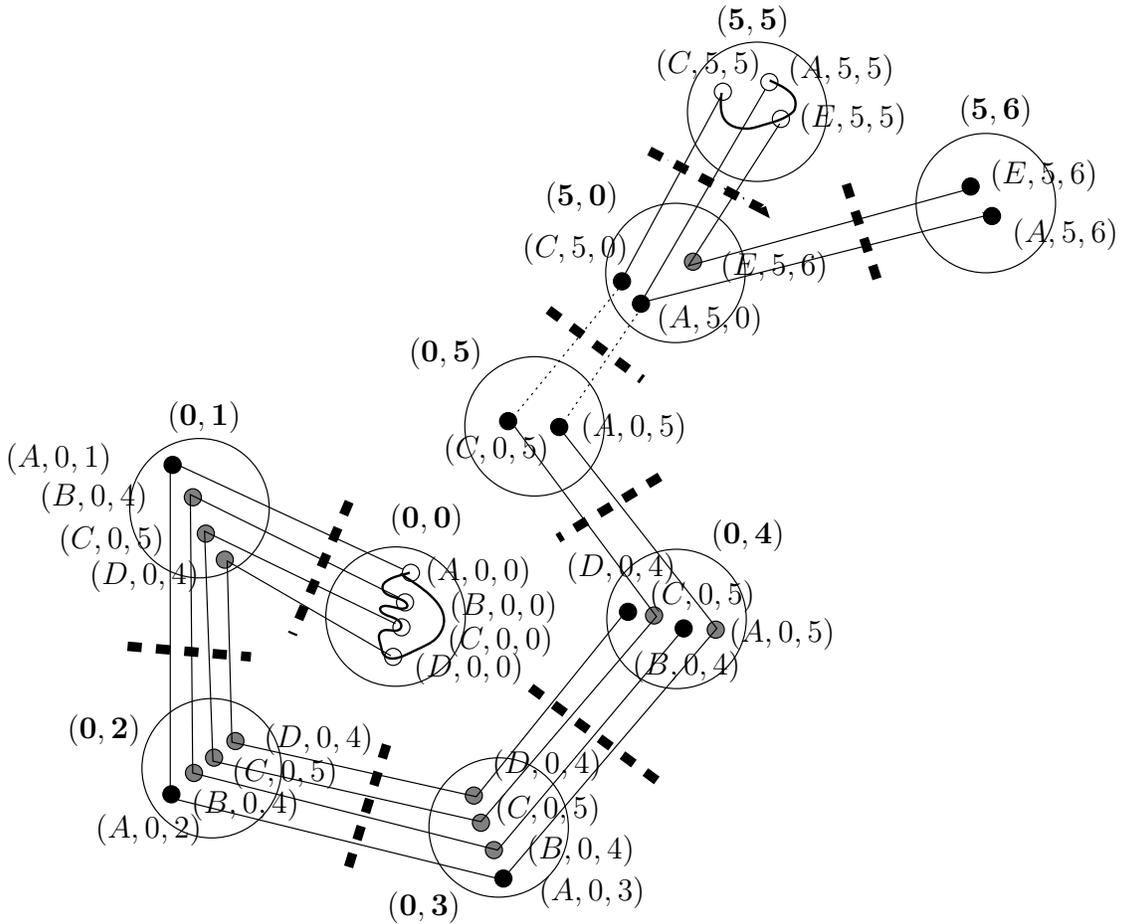}
\caption{Part of the final tree decomposition corresponding to the original
Bags $0$ and $5$ {\em before} the bridging operations,
which will create a sequence of new bag between bags separated by dash bars
in the Figure. The shaded vertices are the re-occurrences of a non-shaded (either black or white) vertices.
The edges between non-shaded vertices will be contained in some are contained in some bag.
The bridging operations will increase the treewidth by $1$.}
\label{fig:new-decomp}
\end{figure}

\begin{claim} The graph $G'$ satisfies $\width(G')\le w+1$.
\end{claim}
\begin{proof}
Our construction of a tree decomposition $\mathcal{T}'$ for $G'$
uses the following procedure. Suppose at some stage we have
constructed two bags represented as ordered sets $ A=\{v_1, v_2,
\cdots, v_t\}$, $ B=\{v'_1, v'_2, \cdots, v'_t\}$. Then {\em
bridging} the two bags involves the following steps:
\begin{enumerate}[(Step 1)]
\item Create bags $ B_1$, $ B_2$, $\cdots$, $ B_{t}$, where $\forall i, 1\le i\le t$,
\[  B_{i}=\{ v_1, v_2, \cdots, v_{i-1}, v_i, v_i', v_{i+1}', \cdots, v_{t}'\}.\]
\item Connect $ A$ to $ B_1$, $ B_i$ to $ B_{i+1}$, $1\le i\le t-1$,
and $ B_t$ to $ B$.
\end{enumerate}
We refer to the constructed path graph as the {\em bridge} bridging
$ A$ and $ B$, and each $ B_i$ as a vertex in the bridge. Note that
for each $i$, $1\le i\le t$, $| B_i|\le t+1$, and $v_i$ or $v_i'$
appear together in $ B_i$. When we apply bridging, vertices in the
two bags will be ordered so that the pair $v_i$ and $v_i'$ take
values $(v, B_1, B_2)$ and $(v, B'_1, B'_2)$ in $V'$ for some common
underlying $v\in V_1$.

We now describe the construction of $\mathcal{T}'$.
\begin{enumerate}[(Step 1)]
\item For each bag $B_0\in \mathcal{V}_1$, adjacent to bags $B_1\prec B_2\prec\cdots\prec B_t$ in $\mathcal{T}_1$,
create bags $(B_0, B_i)$, $\forall i,\ 0\le i\le t$.
For each $v\in B_0$, let $B_{i_1}\prec B_{i_1}\prec\cdots\prec B_{i_\ell}$ be the neighbors
of $B_0$ in $T_v$. Place $(v, B_0, B_0)$ in $(B_0, B_0)$ (thus an inter-tree edge connecting
$(v, B_0, B_0)$ and $(u, B_0, B_0)$ is contained in the bag $(B_0, B_0)$).
For all $j$, $1\le j\le \ell$, place $(v, B_0, B_{i_j})$ in
$(B_0, B_i)$, $\forall i,\ i_{j-1}< i\le i_j$, with $i_0=0$.
In particular, any $(v, B, B')\in V'$ is contained in $(B, B')$.
\item Bridge $(B_0, B_i)$ with $(B_0, B_{i+1})$, $0\le i \le t-1$. Thus
the ordering edge connecting $(v, B_0, B_i)$ and $(v, B_0, B_{i+1})$
is contained in some vertex of this bridge.
\item If $BB'\in \mathcal{E}_1$, bridge $(B, B')$ and $(B', B)$.
Thus an intra-tree edge connecting $(v, B, B')$ and $(v, B', B)$ is
contained in a vertex of this bridge.
\end{enumerate}

Figure~\ref{fig:new-decomp} shows part of the tree decompositions before
the bridging operations for the ternary expansion in Figure~\ref{fig:stage2-exp}.

We have shown above that each edge is contained in some bag. Thus we
need only to show that bags containing a vertex must form a subtree.
By construction, any vertex $(v, B, B)$ is contained in a single bag
$(B, B)$, which forms a tree. Let $(v, B_0, B)\in V'$ and $B_0\ne
B$. Suppose the neighbors of $B_0$ in $\mathcal{T}_1$ are $B_1\prec
B_2\prec \cdots \prec B_t$, and those containing $v$ are
$B_{i_1}\prec B_{i_2}\prec \cdots \prec B_{i_\ell}$. Also,
$B=B_{i_j}$, for some $j$, $1\le j\le \ell$, and let $j_0=
i_{j-1}+1$ (with $i_0=0$). Then $(v, B_0, B)$ is placed in $(B_0,
B_{j_0})$, $(B_0, B_{j_0+1})$, $\cdots$, $(B_0, B_{i_j})$, as well
as all the vertices bridging them together. In addition, $(v, B_0,
B)$ also appears in a connected subgraph of the bridge bridging
$(B_0, B)$ and $(B, B_0)$. Thus the bags containing $(v, B_0, B)$
form a path. This completes the proof that $\mathcal{T}'$ is a tree
decomposition for $G'$.

Recall that $(B, B)$ and $B$ are of the same size, while for $i\ne
0$ $|(B_0, B_i)|\le |B|$. Furthermore, the size of a vertex in a
bridge is at most $1+$ the size of the end bag. Therefore, we
conclude that $\width(\mathcal{T}')\le \width(\mathcal{T}_1)+1
=\width(\mathcal{T})+1=w+1$.
\end{proof}

\section{The treewidth bound in Theorem \ref{th:main} is sharp}
\label{sec:increase}

% We construct a family of graphs $\bar K_{n, n}$ such that for
% sufficiently large $n$, $\width(\bar K_{n, n})=n-1$ but any ternary
% expansion $G_n$ must have $\width(G_n)=n$.
  We now demonstrate a family of graphs of growing treewidth, for
  which any ternary expansion increases treewidth by one.

Let $K_n$ be the complete graph with $n$ vertices ($n$-clique), and
$K_{n, n}$ be the complete bipartite graph whose two sets of
vertices are denoted by $\{ v_i : i\in\mathbb{Z}_n\}$ and $\{ v'_i :
i\in\mathbb{Z}_n\}$. The graph $\bar K_{n, n}$ is obtained from
$K_{n, n}$ by deleting the edges $v_iv'_i$, $i\in \mathbb{Z}_n$.
%Figure~\ref{fig:knn} in the Appendix illustrates $\bar K_{n, n}$.
Recall that for a graph with $2n$ vertices, a {\em perfect matching}
is a subgraph consisting of $n$ vertex-disjoint edges. We call two
edges $e_1, e_2\in E$ {\em engaged} if there is an edge $uv\in E$
such that $e_1$ and $e_2$ are incident to $u$ and $v$, respectively.

\begin{definition} Let $G=(V, E)$ be a graph with $|V|$ even.
A perfect matching $M$ of $G$ is called a {\em bramble matching} if
any two edges in $M$ are engaged in $G$.
\end{definition}

Since contracting edges in a bramble matching of $G=(V, E)$ produces
$K_{|V|/2}$, we immediately obtain

\begin{proposition}
\label{res:cliqueminor} If $G=(V, E)$ admits a bramble matching then
$w(G)\ge |V|/2-1$.
\end{proposition}

\begin{proposition} For $n\ge 3$, $w(\bar K_{n, n})= n-1$.
\end{proposition}

\begin{proof} If $n\ge 3$, $i-1\ne i+1$, for any $i\in\mathbb{Z}_n$.
Thus the edges $v'_{i-1}v_i$ and $v'_iv_{i+1}$ are connected by
the edge $v'_{i-1}v_{i+1}$. Therefore, the edges
$\{ v_iv'_{i+1}: i\in\mathbb{Z}_n\}$ form a bramble matching.
By Proposition~\ref{res:cliqueminor}, $\width(\bar K_{n, n})\ge n-1$.
On the other hand, the following is a tree decomposition of $\bar
K_{n, n}$ with width $n-1$: %(shown in Figure~\ref{fig:knn} in the Appendix)
 a center bag containing $\{v_i:
i\in\mathbb{Z}_n\}$, and $n$ bags $B_k$, $k\in\mathbb{Z}_n$,
connected to it, where $B_k$ includes $\{v_i : i\ne k\}\cup
\{v'_k\}$. Thus $\width(\bar K_{n, n})=n-1$.
\end{proof}

\begin{proposition}
Let $n\ge 19$, and $G_n$ be a ternary expansion of $\bar K_{n, n}$.
Then $\width(G_n)\ge n$.
\end{proposition}

\begin{proof}
For a vertex $v\in V(\bar K_{n, n})$, denote its expansion in $G_n$
by $T_v$ (which is a tree). Contract all vertices  in $T_{v_i}$ and
$T_{v_i'}$ for $i\ne 0$. According to the Tree-partitioning Lemma
\ref{res:treepart} in Appendix B, there must exist an edge $uv$ in
$T_{v_0}$ such that the number of leaves closer to $u$ (than to $v$)
is between $(n-1)/3$ and $2(n-1)/3$. Contract all edges other than
$uw$. Similarly, in $T_{v_0'}$, contract all edges except an edge
$u'w'$ having the same property as $uw$. Denote the resulting graph
$G=(V, E)$. Then each of $\{u, w, u', w'\}$ is connected to $\ge
(n-1)/3\ge 6$ neighbors from the rest of $V$. Without loss of
generality, assume that the following are edges of $G$:
\[ \{ uv'_1, wv'_2, uv'_3, wv'_4, u'v_5, w'v_6, u'v_7, w'v_8\}.\]

\begin{figure}[tbph]
\centering
\includegraphics[width=4.5in]{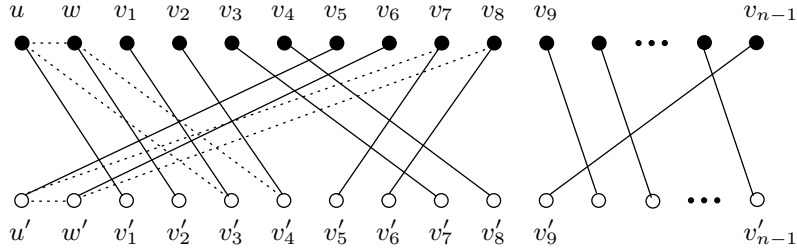}
\caption{The solid lines form the bramble matching $M$, defined in Equation~\ref{eqn:M}.}
\label{fig:bramble}
\end{figure}

We construct a brambling matching $M$ of $n+1$ edges as shown in Figure~\ref{fig:bramble}:
$M=M_1\cup M_1' \cup M_2$, where
\begin{eqnarray}
 M_1&=&\{ uv'_1, wv'_2, u'v_5, w'v_6\},\nonumber\\
M_1'&=&\{ v_1v_3', v_2v'_4, v_3v_7', v_4v_8', v_5'v_7, v_6'v_8\},\quad\textrm{and,}\nonumber\\
M_2&=&\{ v_{i} v'_{i+1} : 9\le i\le n-2\}\cup\{ v'_9v_{n-1}\}.\label{eqn:M}
\end{eqnarray}

By direct inspection, $M$ is a bramble matching. Therefore,
$\width(\bar K_{n, n})\ge n$, by Proposition~\ref{res:cliqueminor}.
\end{proof}

\section{Discussion}
\label{sec:discussion}

  The main result of this paper is good news for many application domains
  and generally means that sparsification of instances does not adversely
  affect their intrinsic complexity, if performed carefully.
  Theorem \ref{th:main} shows that such an expansion does not need
  to increase the treewidth by more than 1.
  For example, extending CNF-SAT with equality clauses (which is trivially
  supported by most SAT solvers today), one can reduce CNF-SAT to 3,3-SAT,
  where each clause has up to three literals and every variable participates
  in at most clauses. Such a reduction could simplify the design and
  implementation of high-performance SAT solvers, which have become
  a popular topic in the last 10 years, thanks to important applications
  in AI, VLSI CAD, logistics, etc. In this special case, our graph-theoretical result
  is consistent with domain-specific work in \cite{ZD07}, which reduces SAT to 3-SAT
  in a different way, but may also increase treewidth by at most one.
  Another application would be to optimize VLSI circuits for better layout
  by breaking down large AND, OR, XOR gates into trees of smaller gates
  and applying fan-out optimization using buffer insertion (as outlined
  in the Introduction).

  The step-by-step description of our algorithm in the proof of
  Theorem \ref{th:main} may seem daunting, but the main insight
  behind this algorithm is rather simple --- to expand a given vertex,
  one must replicate the tree structure of a good tree decomposition around
  this vertex. We hope that the idea to reuse the local structure of tree
  decompositions will find other uses as well. Perhaps, the most
  surprising part of our work is the detailed analysis of how
  treewidth can change during the proposed construction --- it can
  grow only by 1, regardless of the parameters of the input graph,
  and sometimes cannot be preserved by any expansion.

  An interesting direction to extend our main result is to
avoid the reliance of our algorithm on a tree decomposition.
  Perhaps, such an algorithm might also help in constructing a tree decomposition
  or determining the treewidth.
It is also an interesting graph-theoretical problem to characterize
the class of graphs that admit ternary expansions of the same
treewidth. Not containing the graphs in our example as a minor
appears to us a likely characterization.

\bibliographystyle{abbrv}

\pagebreak
\section*{Appendix A: Tree-partitioning Lemma}
\begin{lemma}
\label{res:treepart}
   Consider a tree $T$ with $k\geq 3$ leaves in which no vertex has degree
   greater than three. While every tree-edge separates the tree into
   two connected components, for at least one edge each component
   contains between $ k/3 $ and $ 2k/3-1 $ leaves, inclusively.
\end{lemma}
\begin{proof}
We regard $T$ as a tree rooted at a non-leaf vertex $r$.
Note that each vertex, except $r$, has no more than two children.
For a vertex $v$ of $T$, define the size of $v$, $s(v)$,
to be the number of leaves of the subtree rooted at $v$.
Let $v$ be the child of $r$ with the largest size among its siblings.
Then $s(v)\ge k/3$. If $s(v)< 2k/3$, then $rv$ satisfies the requirement.
Otherwise, traverse down the tree following
the edge that connects the child with the larger size.
Suppose $uv$ is the edge so that $v$ is the first vertex
encountered having $< 2k/3$ size.
Since $s(u)\ge 2k/3$, and $s(v)\ge s(u)/2 \ge k/3$,
$uv$ satisfies the requirement.
\end{proof}

%\newpage
\section*{Appendix B: Figures}

\begin{figure}[tbph]
\centering
\includegraphics[width=6.2in]{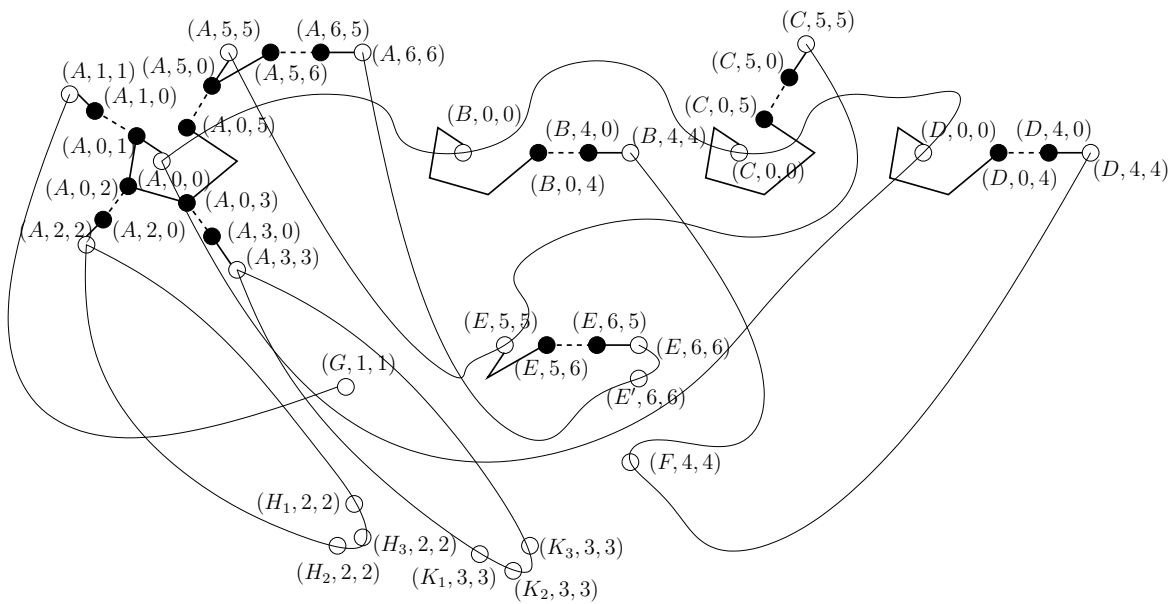}
\caption{The ternary expansion of the graph in Figure~\ref{fig:stage2} after
the steps in Stage 2. The straight solid lines are {\em ordering} edges,
the dashed lines are {\em intra-tree} edges, and the curves are {\em inter-tree} edges.
The expansions of vertex $A$, $B$, $C$, and $D$ are drawn in separate
positions at the upper part, and all the others are at the lower part.}
\label{fig:stage2-exp}
\end{figure}

%\clearpage

% \begin{figure}[tbph]
% \centering
% \includegraphics[width=5in]{knn}
% \caption{The graph $\bar K_{n, n}$ and its tree decomposition with width $n-1$.}
% \label{fig:knn}
% \end{figure}

\end{document}